\documentstyle[aasms4]{article}
\begin{document}

\title{Complete Zeldovich approximation}

\author{J. Betancort-Rijo, M. L\'opez-Corredoira}
\affil{Instituto de Astrof\'{\i}sica de Canarias, E-38200 La Laguna,
Tenerife, Spain}

\begin{abstract}
 We have developed a generalization of the Zeldovich approximation (ZA) that is
 exact in a wide variety of situations, including plannar, spherical and
 cilyndrical symmetries. We have shown that this generalization, that we call
 complete Zeldovich approximation (CZA), is exact to second order at an arbitrary 
 point within any field. For gaussian fields, the third order error have been
 obtained and shown to be very small. For statistical purposes, the CZA leads to
 results exact to the third order.
\end{abstract}
 
 
 Developing a simple analytical approximation that accounts accurately for the
 non-linear evolution of density fields seems to be a rather interesting
 task. To reconstruct with some accuracy the initial conditions from
 present velocity field in the line of sight 
 one such approximation is needed. To obtain the
 statistical properties of the present field in terms of those for the initial
 one; the non-linear corrections to the microwave background or to the
 Gun-Peterson effect, an approximation of this kind is highly convenient.
 An approximation that has been widely used to this purposes is Zeldovich
 approximation (ZA), where the density fluctuation, $\delta $, is a unique function of
 the proper values, $\lambda _i$, of the linearly calculated local deformation tensor
 $\frac{\partial \vec{u}}{\partial \vec{x}}$ ($\vec{u}$
 being the peculiar velocity field).
 
 \begin{equation}
 (1+\delta )^{-1}=\prod _{i=1}^3 (1-\lambda _i)
 \label{1}
 \end{equation}

 This approximation give generaly rather good results and it is particularly
 convenient for deriving the statistical properties of the present field, since 
 we only need the statistical properties of the $\lambda _j$ in the initial field.
 In this approximation the local deformation tensor is that given by the linear
 theory. So, although it is a good approximation to all orders, it is exact only
 to first order. On the other hand, in the Lagrangian perturbative development
 (LPD) the deformation tensor is formaly exact (Bouchet et al. 1995), but it is
 not a unique function of the $\lambda _i$. 
 Within this context, the question easily arise
 as to whether is posible to find an approximation depending only on the $\lambda _i$ and
 substancialy more accurate than ZA.
 Reisnegger \& Miralda-Escude (1995) considered an extension of ZA (EZA) that is exact
 for planar, spherical and cilyndrical symmetry. However, although this
 approximation gives usually better results than ZA, it is not fully consistent.
 Hui and Bertschinger (1996) developed an approximation (LTA) that is exact for 
 any fluid element such that the orientation and axis ratios of the
 gravitational and velocity equipotentials are constant along its trayectory. This
 include planar, spherical and cilyndrical symmetries. The problem with this
 approximation, which gives excelent results, is its complexity, that makes very
 difficult to determine the explicit dependence on the $\lambda _i$.
 The purpose of this letter is to present the explicit dependence of $\delta $ as a
 function of the $\lambda _i$ for the most accurate approximation that is a unique 
 function of these quantities. This approximation we call complete Zeldovich
 approximation (CZA). Like EZA and LTA, CZA is exact for planar ($\lambda _1\ne 0$,
 $\lambda _2=\lambda _3=0$), 
 spherical ($\lambda _1=\lambda _2=\lambda _3$)
 and cilyndrical ($\lambda _1=\lambda _2$, $\lambda _3=0$)
 symmetries, but, unlike those approximations, it is
 also exact in another wide variety of cases that
 essentialy correspond to the collapse of an initial top-hat elliptical density
 fluctuation. We have shown that CZA is exact to second order and computed the
 third order error, which is very small. We have also shown that the exact
 evolution at an arbitrary point may be expressed in terms of a simple extension
 of the CZA containing some aditional variables (the CZA being recovered when
 these variables are set equal to zero). All this is shown in detail in an
 accompanying paper; here we shall only present the CZA and comment upon its
 meaning and derivation.
 
 Approximations depending only on the $\lambda _i$ are usually called local.
 We shall retain this convention, but it must be noted that the $\lambda _i$,
 although defined locally, are non-locally generated. The sum of the $\lambda _i$,
 which is equal to the linear density perturbation, $\delta _L$, 
 can take any value at a given point, regardless of the values
 taken at other points. To obtain the $\lambda _i$, however,
 we must integrate the continuity
 equation, so the $\lambda _i$ at a given point depends on the whole field 
 $\delta _L(\vec{x})$. Another way to
 point out the non-local caracter of the $\lambda _i$ is 
 by noticing that the quantities $\lambda _i-\frac{\delta _L}{3}$ 
 are generated by the action of the linearly calculated local tidal field which, obviously, 
 depends on the whole field $\delta _L(\vec{x})$. Keeping this in mind, it is not
 surprising that although knowing $\delta _L$ at a point at some initial time let us
 to obtain the evolution only to first order, knowing the $\lambda _i$ let us to
 obtain the evolution to second order.
 We shall now describe the steps we have followed to obtain the CZA. We choose
 the ansatz:
 
 \begin{equation}
 (1+\delta )^{-1}=\prod _{i=1}^3(1-r_i(\lambda )\lambda _i)
 \label{2}
 ,\end{equation}
 where $r_i(\lambda )$ are certain functions of the $\lambda _i$.
 Within this ansatz, ZA corresponds to the
 zeroth order approximation for $r_i$ ($r_i=1$). It is interesting to note that,
 although independently developed this ansatz is similar to that chosen by the
 authors of the EZA, except for the fact that they assumed all $r_i$ to be equal; an
 assumption that is incompatible with exactness to second order.
 
 To determine the functions $r_i$ we use the constraints imposed on them by:
 considerations about the symmetry of $\lambda _i$ with respect to permutations of the
 indexes; the fact that for planar symmetry (\ref{2}) must be exact with $r_i=1$;
 the form of the exact solution for  spherical collapse and compatibility of the form
 of $r_i$ with dynamical equations. These conditions determine $r_i$ uniquely. Let us
 comment them in more detail.
 
 Rotational invariance imply that the $r_i$ must reduce to each other through
 permutations of the indexes. So, they all derive from the same function, $r(\vec{u})$.
 
 \begin{equation}
 r_i(\vec{\lambda })=r(\vec{u})|_{\vec{u}=(\lambda _{i}, \lambda _{j},
 \lambda _{k})}
 \label{3}
 \end{equation}

 Furthermore, rotational symmetry in the plane perpendicular to the i-th proper
 axis imply that $\lambda _j$, $\lambda _k$ must enter symmetricaly in this expression.
 So, $r(\vec{u})$ must be symmetric with respect to its second and third arguments.
 We now assume that a series expansion of $r(\vec{u})$ in powers of the $u_i$ exist.
 We shall see later that this series converges for all relevant $\vec{u}$. 
 The symmetry considerations we have
 just mentioned, imply that this series can only contain terms of the form:
 
 \[
 r(u_1, u_2, u_3)=1+\sum _{l,m,n=0}^\infty C_{l,m,n}^p
 (u_2+u_3)^l(u_2-u_3)^{2n}u_1^m
 \]\begin{equation}
 p\equiv l+2n+m
 \label{4}
 ,\end{equation}
 where $C_{l,m,n}^p$ are the coefficients of the $p$-th order terms.
 Noting that in the
 planar case ($\lambda _1\ne 0$, $\lambda _2=\lambda _3=0$) expression (\ref{2}) 
 is exact with $r_i=1$, it is
 clear that expansion (\ref{4}) cannot contain terms with $m\ne 0$ and $l=n=0$. 
 So, in our notation we must have:
 
 \begin{equation}
 C_{0,m,0}^p=0
 \label{5}
 \end{equation}
 
 Hence ,there are $2p-1$ terms of order $p$.
 
 For spherical collapse, both the actual density fluctuation, $\delta _{sp}$, 
 and its linear value, $\delta _L$, may be expressed exactly as a 
 parametric function of time (Peebles 1980). From these expressions we have derived 
 an expression for $\delta $ as an explicit function of $\delta _L$:
 
 \[
 1+\delta _{sp}=\left(1-r_{sp}(\delta _L)\frac{\delta _L}{3}\right)^{-3}
 \]\[
 r_{sp}(\delta _L)=1+f_1(\theta )\frac{\delta _L}{7}+
 f_2(\theta )\frac{23}{567}\delta _L^2+f_3(\theta )\frac{13}{900}\delta _L^3
 +f_4(\theta )5.86\times 10^{-3}\delta _L^4+f_5(\theta )2.55\times 10^{-3}
 \delta _L^5+R_{sp}(\delta _L)
 \]\[
 R_{sp}(\delta _L)=f_6(\theta )2.58\times 10^{-3}\delta _L^5\left(
 \frac{1}{1-\frac{\delta _L}{2.065}}-1\right)+E
 \]\begin{equation}
 |E|<2\times 10^{-3}\ {\rm for \ } \delta _L\le 1.57; \ \ E \propto O(\delta _L^6)
 \label{6}
 ,\end{equation} 
 
 where $\theta $ stands for all cosmological parameters. For a flat  Friedman model all 
 $f_i(\theta )$ are exactly equal to one. For a general Friedman model the dependence on 
 $\Omega $ (the density in units of the critical one) is very mild. When $\Omega >1/20$ 
 the following is a rather good approximation:
 
 \begin{equation}
 f_i(\Omega )=\Omega ^{2i/63}
 \label{7}
 \end{equation}
 
 Comparing expressions (\ref{6}) and (\ref{2}) and noting that for spherical symmetry       
$\lambda _1=\lambda _2=\lambda _3=\frac{\delta _L}{3}$, 
we obtain the following constraint on $r(u)$:
    
   \begin{equation}
   r_{sp}(\delta _L)=r\left(\frac{\delta _L}{3},\frac{\delta _L}{3},\frac{\delta _L}{3}\right)
\label{8}
   \end{equation}

    This relationship imply that the coefficients of order $p$, $C_{l,m,n}^p$, must satisfy
    just one equation. So ,for $p$ larger than one the coefficients are
    underdetermined. However, the value of $\delta $ given by (\ref{2}) must satisfy the
    dynamical equations (Peebles 1980):
    
    \[
    \frac{d\vec{v}}{dt}+\vec{v}\frac{\dot{a}}{a}=-\frac{\nabla \phi }{a}; \ 
    \phi (\vec{x})=G\ a^2p_b(\tau )\int \frac{d^3x'\delta (x')}{|x'-x|}
    \]\begin{equation}
    \vec{v}=a\vec{u}; \ \vec{u}\equiv \dot{\vec{x}}; r_i(\lambda )\lambda _i=
    -\frac{\partial \dot{x_i}(t, q_i)}{\partial q_i}
    \label{9}
    \end{equation}
    
    where $x_i$ are Eulerian comoving coordinates and $x_i (t,q_i)$ are the Eulerian
    coodinates of a particle with Lagrangian coordinates $q$ (the last
    relationship holds in the local proper system). Note that the continuity equation
    is automaticaly satisfied by CZA.
    From these equations one may see that terms of order $p$ in $r$ imply
    the existence of certain terms of order $p+1$. This recursive scheme, together
    with expresion (\ref{8}) determine completely all coefficients. Their computation,
    which is not trivial, is given in detail in the accompanying paper. Here we
    simply give the result and comment it.
    It is interesting to note that the process we have followed to determine the
    coefficients essentialy amounts to analyticaly continue the $r_i(\vec{\lambda })$ 
    known in the planar and spherical case in a manner consistent with equations (\ref{9}).
    The situations described exactly by the CZA are those where the local
    deformation tensor (whose proper values are $r_i(\vec{\lambda })\lambda _i$) are everywhere the
    same (or, at least at all fluid elements affecting each other's
    evolution). This must be so because it is true for the planar and spherical
    case and preserved by the continuing procedure.
    For a flat Friedman universe, we have found for $r_i$ (in the general
    case terms of order $p$ should be multiplied by $f_p(\theta )$):
    
    \[
     r_i(\lambda _i, \lambda _j, \lambda _k)=1+\frac{3}{14}
     (\lambda _j+\lambda _k)+\frac{18}{245}(\lambda _j+\lambda _k)^2+
     \frac{157}{4410}(\lambda _j+\lambda _k)\lambda _i
     +\frac{3}{245}(\lambda _j-\lambda _k)^2+0.03371(\lambda _j+\lambda _k)^3
     \]\[
     +1.63\times 10^{-2}(\lambda _j+\lambda _k)^2\lambda _i
     +2.75\times 10^{-2}(\lambda _j+\lambda _k)\lambda _i^2
     +10^{-3}(\lambda _j-\lambda _k)^2\lambda _i
     +1.2\times 10^{-2}(\lambda _j-\lambda _k)^2(\lambda _j+\lambda _k)
     \]\[+
     1.94\times 10^{-2}(\lambda _j+\lambda _k)^4
     +9.4\times 10^{-3}(\lambda _j+\lambda _k)^3\lambda _i
     +1.58\times 10^{-2}(\lambda _j+\lambda _k)^2\lambda _i^2
     +1.3\times 10^{-2}(\lambda _j+\lambda _k)\lambda _i^3
     \]\begin{equation}+
     4.3\times 10^{-3}(\lambda _j-\lambda _k)^4
     +8.4\times 10^{-3}(\lambda _j-\lambda _k)^2(\lambda _j+\lambda _k)^2
     +7.2\times 10^{-4}(\lambda _j-\lambda _k)^2(\lambda _j+\lambda _k)\lambda _i
+R(\lambda _i, \lambda _j, \lambda _k)
\label{10}
\end{equation}

    This messy expression includes explicitly terms up to the fourth  order
    although for more purposes using up to the quadratic term is enough. Terms of
    order larger than four are usualy very small. However, in the rare cases when
    they are of some relevance many orders contribute roughly equaly. So, it is
    not convenient to include higher order terms explicitly. Instead, we
    approximate all those terms by a single term, $R$, that we shall latter give.
     Expression (\ref{2}) with $r_i$ given by (\ref{10}) 
     give the exact evolution of $\delta $ in a field
     where the local deformation tensor is everywhere given by  
     $r_i(\vec{\lambda })\lambda _i$    at a time when the linear one
     is $\lambda _i$. It is clear that this is the situation within a top-hat cilyndrical
     fluctuation. So, we may use this case (with $\Omega =1$) to check the 
     correctness of the 
     continuing process. In this case we have $\lambda _1=0$, $\lambda _2=\lambda _3=
     \frac{\delta _L}{2}$. So,
     expressions (\ref{2}) and (\ref{10}) lead to:
     
     \[
     (1+\delta _{\rm cyl})=\left(1-r_{\rm cyl}(\delta _L)\frac{\delta _L}{2}\right)^2; \ 
     r_{\rm cyl}\equiv r\left(\frac{\delta _L}{2},\frac{\delta _L}{2},0\right)
     \]\begin{equation}
     =1+\frac{3}{28}\delta _L+\frac{107}{3528}\delta _L^2+
     1.135\times 10^{-2}\delta _L^3+4.5\times 10^{-3}\delta _L^4+
     R_{\rm cyl}(\delta _L)
     \label{12}
     \end{equation}

     On the other hand, we have obtained $r_{\rm cyl}(\delta _L)$ 
     through direct accurate numerical integration (the
     error of $r_{\rm cyl}<10^{-5}$) and fitted the coefficients in the expansion.
     These cofficients agree with the predictions (those in (\ref{12})) 
     well within the fitting
       errors (0.1\% ,0.4\% ,3\% ,10\% for coefficients from the first to the
       fourth). We have also used the numerical results to fit an approximate
       expression for $R_{\rm cyl}(\delta _L)$ and find:
       
\[
R_{\rm cyl}=2.2\times 10^{-3}\delta _L^5\left(1-\frac{\delta _L}{2.06}\right)^{-1}
+E
\]\begin{equation}
|E|<5\times 10^{-3}
\label{13}
\end{equation}     
       
       We may now obtain an expression for $R(\lambda )$ (see expression (\ref{10})) demanding
       that the exact result be obtained in the planar and spherical cases and 
       that it reduces to a good approximation to $R_{\rm cyl}$ 
       in the cilyndrical case:
       
\[
R(\lambda _i, \lambda _j, \lambda _k)=\left[1-9\left(\lambda _i-\frac{\lambda _j+
\lambda _k}{2}\right)\left(1-\frac{\lambda _i+\lambda _j+\lambda _k}{1.3}\right)\right]
\]\[ \times  
\left(V_{\rm sp}(\lambda _i+\lambda _j+\lambda _k)-V_{\rm sp}(\lambda _i)+
V_{\rm sp}\left(\frac{\lambda _j+\lambda _k}{2}\right)\right)+E
\]\[
V_{\rm sp}(x)\equiv r_{\rm sp}(x)-\left(1+\frac{x}{7}+\frac{23}{567}x^2+
\frac{13}{900}x^3+5.86\times 10^{-3}x^4\right)
\]\begin{equation}
=2.58 \times 10^{-3}x^5\left(1-\frac{x}{2.06}\right)^{-1}+E_2; \ 
E_2<2.3\times 10^{-3} \ \ {\rm for }\ x<1.57
\label{14}
\end{equation}

       This expression corresponds to $\Omega =1$ and may be inmediately generalized
       for an arbitrary cosmological model. The maximum error of expression 
       (\ref{10}) with $R(\lambda )$ given by (\ref{14}) is $6\times 10^{-3}$ being usualy quite smaller.
       For general values of the $\lambda _i$
   the situation described exactly by expressions (\ref{2}) and (\ref{10}) 
   is more complex than for the three peculier cases already
       considered. However, for all practical purposes a simple generalization of
       these cases, namely, a top-hat initial ellipsoidal fluctuation, may be
       considered to be described exactly by those expressions. In fact, it may
       be shown that the intrinsic error of $r_i(\vec{\lambda })$ 
       in these situations 
       as given by (\ref{10}) ($\approx 3\times 10^{-3}$ at the time of collapse) 
       is smaller than the error of this expression (due only to $R(\lambda )$).
       
       In the top-hat spherical and cilyndrical cases the deformation tensor for
       outside matter is diferent from that for matter within. The same is true
       for the top-hat eliptical case. However, in this case, unlike in the
       former the outside matter is relevant to the evolution of the matter
       within. Expression (\ref{10}) accounts exactly to the second order (and a small
       error to higher orders) for the small contribution of the outside matter
       to the tidal field within the ellipsoid.
        
      To check the accuracy of the LTA approximation, Hui and Bertschinger (1996) considered
      the collapse of a top-hat initial fluctuation with axial ratios 1:1.25:1.5
      and represented (in their fig.2 ) the evolution of the axis predicted by
      this approximation and that predicted by an approximation (that they
      called exact) that neglects the effect of outside matter. The ellipsoid
      generates a linear growing mode for the velocity field with asociated 
      values given by: $\lambda _1=0.2576a$; $\lambda _2=0.3233a$; 
      $\lambda _3=0.4191a$. (label ``1'' corresponding to the
      largest axis). As we have said before, expressions (\ref{2}) and (\ref{10}) may be
      considered exact in this case, giving for the evolution of the axis:
      
      \[x_i=w_ia(1-r_i(\vec{\lambda })\lambda _i)
      \]
      
      where $a$ is the expansion factor of the universe, and $w_i$ are the axis ratios.
      This is represented in figure 1 along with the predictions of an
      approximation where all $r_i$ are set equal to their symmetrized value 
      $\left(\frac{r_1(\lambda )+r_2(\lambda )+r_3(\lambda )}{3}\right)$. 
      This approximation must be very close to EZA (Reisenegger and
	    Miralda-Escude 1995). It may be shown that by symmetrizing the
	    effect of the outside matter is neglected. This may be checked by
	    noting that this approximation is barely distinguishable from 
	    ``exact''.
	    The LTA and CZA are indistinguishable (within a 0.2\% ) up to 
	    a $\sim 1.2$, differing very little up to the collapse, that takes place at
	    $a=1.584$ for CZA and $a=1.613$ for LTA. Note that the evolution of the
	    difference between the values of the axes given by these
	    approximations is quantitatively the same as that between the  exact
	    solution and the LTA for an elliptical cloud in empty space (see
	    their fig. 3), such as would be expected if CZA were exact. In fact,
	    it may be shown that the error of the value of a at colapse given by
	    CZA is at most 0.3\% .
	    The agreement between LTA and CZA for $a<1.2$ is so complete that if
	    we had an explicit expression (like (\ref{10})) for the 
	    former approximation, the
	    first order coefficient should be very close (within   3\%) to that
	    for the CZA and those of second order should not differ much. This
	    imply that most likely LTA is exact to second order and that it
	    accounts for the effect of outside matter (both facts are
	    related), hence being more accurate than ``exact'' and ECZ (which can
	    not be exact to second order).
	    
	    So far we have considered the situations described exactly by CZA.
	    However we are mostly interested in the performance of this
	    approximation at an arbitrary point. There is no obvious reason to
	    expect an approximation determined by the above considerations to be
	    the best at a random point. To see that this is actualy so we first
	    write  $\delta $ at such point in the form:
	    
	    \[
	    1+\delta =\prod_i \left(1-r_i\lambda _i+\frac{3}{14}\delta _Lx_i+
	    8.46\times 10^{-2}\delta _L^2x_i+8.34\times 10^{-2}
	    \left[\frac{x_1^2+x_2^2+x_3^2}{3}\right]\delta _L+...\right)^{-1}
	    \]\begin{equation}
	    \approx 1+\sum _i\lambda _i+\frac{10}{7}(\lambda _1\lambda _2+...)
	    +\left[\frac{3}{14}\delta _L+8.46\times 10^{-2}\delta _L^2\right]
	    \sum x_i
	    \label{15}
	    \end{equation}
	     where the $x_i$ are certain variables defined by the action of some
	    integral operator on $\delta _L(\vec{x})$ and that cannot be reduced to 
	    functions of
	    the $\lambda _i$. This expression, that in full would contain 
	    more variables,
	    is formaly exact, like the LPD (Bouchet et al 1995). 
	    But here for each order we have separated the part depending on the $\lambda _i$, that goes
	    into $r_i(\lambda )\lambda _i$. We may use (\ref{15}) 
	    and the probability distribution of
	    the  $\lambda _i$ at points with a fixed value of $\delta _L$ 
	    to obtain the statistical
	    properties of $\delta $ at these points. By comparing what we find with
	    the exact results found by Bernardeau (1994) for these
	    statistical quantities we find:
	    
	    \[\sum x_i=0; \ \ \langle x_i \rangle _{\lambda }=C\left(\lambda _i-
	    \frac{\delta _L}{3}\right); \ \langle x_i^2\rangle _{\lambda }=
	     \langle x_i\rangle ^2_{\lambda }+\frac{2C}{9}(1-C)\sigma ^2
	     \]\begin{equation}
	     C=\frac{3}{2}6.4(C_{\rm exp}-0.0544)
	     \label{16}
	     \end{equation}

	    The first result is valid at every point and could be derived in
	    other ways ,for example, by comparing (\ref{15}) with the LPD. The other
	    results, which are of statistical character, give the mean and mean
	    quadratic value of $x_i$ over points with a fixed value of the $\lambda _i$. 
	    $C_{\rm exp}$ is a spectral constant defined in the last reference. 

In the context of expression (\ref{15}) the CZA, as we have defined it,
may be caracterized by the neglect of the $x_i$. 
This approximation is the same for
all fields. However, we could consider a CZA specific for each spectrum (for
Gaussian fields) simply by inserting in (\ref{15}) in the place of $x_i$ , 
$x_i^2$, their mean values given in (\ref{16}) with 
the value of $C_{\rm exp}$ corresponding to that spectrum.
This approximation give exactly to third order the moments of $\delta $
over points with fixed $\lambda _i$ values. Hence, it gives the one point 
statistics exactly to third order.
For smooth field (the case considered here) $C_{\rm exp}$ lies 
between 0.053 and 0.061 for most interesting spectrums. So the general 
CZA (which is exact to third order
for $C_{\rm exp}=0.0544$) imply a very small error to third order.
We have stimated the error of the CZA by computing to third order (the first
non-vanishing) the RMS fluctuation of the
value of  $\delta $ over points with fixed $\lambda _i$ values. We found:

\[
\langle (\delta -\langle \delta \rangle_\lambda)^2\rangle _\lambda^{1/2}=
\left[\left(\frac{3}{14}\right)^2(\lambda _1^2+\lambda _2^2+
\lambda _3^2)-(\lambda _1\lambda _2+...)\frac{1}{\gamma \sigma
^2}+3.1\right]^{1/2}\gamma \sigma ^2\delta _L
\]\[
\gamma\le \frac{30}{13}(C_{exp}-0.0471)
\]

The fact that at every point the sum of the $x_i$ vanishes imply (see (\ref{15}))
that CZA is exact to second order. As we have seen, this is most likely to be 
also true for the LTA, but it is not true for the EZA.

Both the ZA and the CZA are unique functions of the $\lambda _i$ so, 
one might wonder where is what makes the latter exact to second order. 
The answer is that in the CZA we
use for the proper values of the deformation tensor
$r_i(\vec{\lambda })\lambda _i$ which is exact to
second order, rather than $\lambda _i$. 
It must be noted however that the velocity field is
not given to second order in an explicit manner.
The equation $\nabla _q\vec{u}=-\sum _ir_i\lambda _i$
is exact to second order but, to obtain the velocity field,
we must integrate it.

\eject

{\bf Figure captions:}

Figure 1: Evolution of the axis lengths for a homogeneous ellipsoid
embedded in an expanding Universe. Initial axial ratios are 1:1.25:1.50.
The solid line corresponds to CZA, which is practically exact;
and the dashed line corresponds to an approximation that makes all
$r_i$ equal to their symmetrized value.

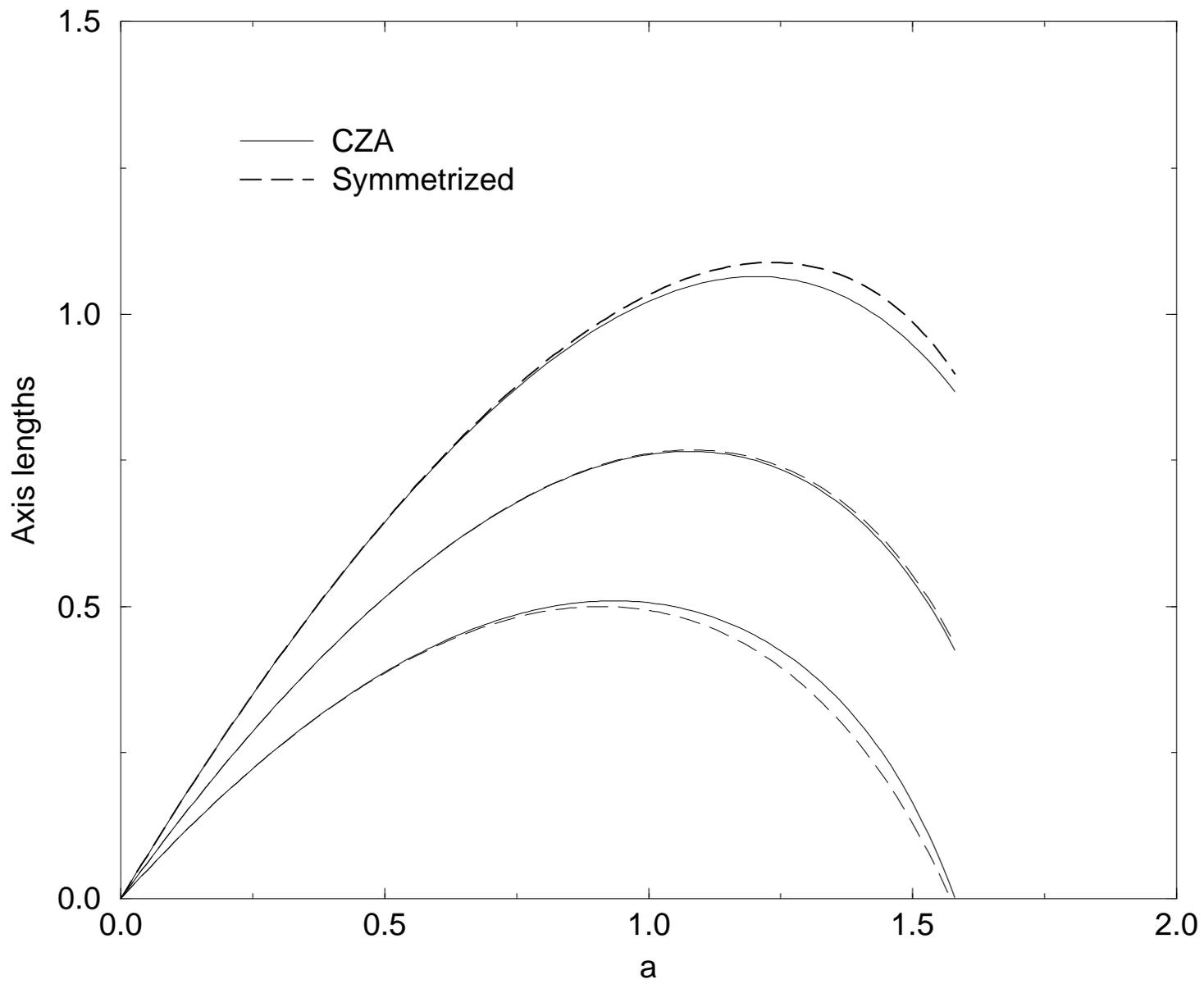
\end{document}